\documentclass[12pt]{article}
\usepackage{amsmath,url}
\usepackage{graphicx}
\usepackage{cite}
\input{epsf}
\providecommand{\href}[2]{#2}

\newcommand\as{\alpha_{\mathrm{S}}}

\def\to{\rightarrow}

\begin{document} 
\renewcommand{\thefootnote}{\fnsymbol{footnote}}
\vspace*{2cm}

\begin{center}
{\Large \bf Vector-boson pair production at NNLO\footnote{Based on talks given at the XLIXth Rencontres de Moriond, QCD and High energy interactions, La Thuile, march 2014, and at Loops and Legs in Quantum Field Theory, Weimar, april 2014.}}
\end{center}

\par \vspace{2mm}
\begin{center}
{\bf Massimiliano Grazzini\footnote{On leave of absence from INFN, Sezione di Firenze, Sesto Fiorentino, Florence, Italy.}}
\vspace{5mm}

Physik-Institut, Universit\"at Z\"urich, CH-8057 Z\"urich, Switzerland

\vspace{5mm}

\end{center}

\par \vspace{2mm}
\begin{center} {\large \bf Abstract} \end{center}
\begin{quote}
\pretolerance 10000

We consider the inclusive production of vector-boson pairs in hadron collisions.
We review the theoretical progress in the computation of
radiative corrections to this process
up to next-to-next-to-leading order in QCD perturbation theory.

\end{quote}

The production of vector-boson pairs is a relevant process
for physics studies within and beyond the Standard Model (SM).
First of all, this process can be used to measure the vector boson trilinear couplings.
Any deviation from the pattern predicted by $SU(2)\otimes U(1)$ gauge invariance would be a signal of new physics.
The Tevatron collaborations have measured $WW$, $ZZ$, $WZ$, $Z\gamma$ and $W\gamma$ cross sections at invariant masses larger than those probed at LEP2,
setting limits on the corresponding anomalous couplings,
and the LHC experiments are now continuing this research program \cite{Wang:2014uea}.
Furthermore, vector boson pairs are an important background for new physics searches.
Although the recently discovered Higgs resonance is well below the $WW$ and $ZZ$ threshold,
the off-shell $WW$ and $ZZ$ backgrounds are crucial both in the extraction of the Higgs signal and in
a measurement of the Higgs boson width \cite{Kauer:2012hd,Caola:2013yja,Campbell:2013wga}.
Possible charged Higgs bosons from non standard Higgs sectors could decay into $WZ$ final states.
Typical signals of supersymmetry, e.g. three charged leptons plus missing energy,
receive an important background in $WZ$ and $W\gamma$ production.
In this contribution we review the current status of theoretical predictions for vector-boson pair production, with emphasis on QCD radiative corrections, and
focusing on NNLO QCD effects in $Z\gamma$ \cite{Grazzini:2013bna}, $W\gamma$ \cite{inprep} and $ZZ$ \cite{Cascioli:2014yka} production (NNLO corrections to $\gamma\gamma$ production have been presented in Ref.~\cite{Catani:2011qz}).

The theoretical efforts for a precise prediction of vector-boson pair production in the SM
started more than 20 years ago, with the first NLO QCD calculations
\cite{Ohnemus:1990za,Mele:1990bq,Ohnemus:1991kk,Frixione:1993yp,Ohnemus:1992jn,Ohnemus:1991gb,Frixione:1992pj} with stable vector bosons.
The computation of the relevant one-loop helicity amplitudes \cite{Dixon:1998py} allowed
complete NLO calculations \cite{Campbell:1999ah,Dixon:1999di,DeFlorian:2000sg} including the leptonic decay,
spin correlations and off-shell effects.
In the case of $WW$, $ZZ$ and $Z\gamma$ production
the loop-induced gluon fusion contribution, which is formally next-to-next-to-leading order (NNLO),
has been computed in Refs.~\cite{Dicus:1987dj,Glover:1988fe,Glover:1988rg,Ametller:1985di,vanderBij:1988fb}.
NLO predictions for vector-boson pair
production including the gluon-induced contribution,
the leptonic decay of the vector boson
with spin correlations and off-shell effects have
been presented in Ref.~\cite{Campbell:2011bn}.
Electro-weak corrections to vector boson pair production
have been considered in Refs.~\cite{Hollik:2004tm,Accomando:2005ra,Bierweiler:2012kw,Bierweiler:2013dja,Baglio:2013toa,Billoni:2013aba}.

The NNLO QCD computation of $VV^\prime$ production requires the evaluation of the tree-level
scattering amplitudes with two additional (unresolved) partons, of the one-loop amplitudes with one additional parton, and of the one-loop-squared and
two-loop corrections to the Born subprocess $q{\bar q}\to VV^\prime$.
Up to now, the bottleneck for the NNLO calculation has been the knowledge of
the relevant two-loop amplitudes.
The two-loop helicity amplitudes for $W\gamma$ and $Z\gamma$ production have been presented in Ref.\cite{Gehrmann:2011ab}.
Recently, a major step forward has been carried out, with the evaluation of all the two-loop  planar \cite{Gehrmann:2013cxs,Henn:2014lfa}
and non planar \cite{Gehrmann:2014bfa,Caola:2014lpa} master integrals relevant
for the production of off-shell vector boson pairs, and the
calculation of the corresponding helicity amplitudes is now feasible.

Even having all the relevant amplitudes, the computation of the NNLO corrections
is still a non-trivial task, due to the presence of infrared~(IR) singularities at
intermediate stages of the calculation that prevent a straightforward application of numerical techniques.
To handle and cancel these singularities at NNLO the $q_T$ subtraction formalism \cite{Catani:2007vq} is particularly suitable, since it is fully developed \cite{Catani:2013tia}
to work in the hadronic production of heavy colourless final states.

In the following we present a selection of numerical results for $Z\gamma$ \cite{Grazzini:2013bna}, $W\gamma$ \cite{inprep} and $ZZ$ \cite{Cascioli:2014yka} production at the LHC.
In the above applications the required tree-level and one-loop amplitudes were
obtained with the
{\sc OpenLoops}~\cite{Cascioli:2011va} generator, which employs the Denner-Dittmaier algorithm \cite{Denner:2005nn} for the numerical evaluation of one-loop integrals and implements
a fast numerical recursion 
for the calculation of NLO scattering amplitudes within the SM.

We use the MSTW 2008 \cite{Martin:2009iq} sets of parton distributions, with
densities and $\as$ evaluated at each corresponding order
(i.e., we use $(n+1)$-loop $\as$ at N$^n$LO, with $n=0,1,2$),
and we consider $N_f=5$ massless quarks/antiquarks and gluons in 
the initial state. As for the electroweak couplings, we use the so called $G_\mu$ scheme,
where the input parameters are $G_F$, $m_W$, $m_Z$. In particular we 
use the values
$G_F = 1.16639\times 10^{-5}$~GeV$^{-2}$, $m_W=80.398$ GeV, $\Gamma_W=2.1054$ GeV, $m_Z = 91.1876$~GeV, $\Gamma_Z=2.4952$~GeV. For simplicity, flavour mixing is neglected, and the CKM matrix is taken to be the unit matrix.

When considering the $V\gamma$ final state ($V=W,Z$),
besides the {\it direct} production in the hard subprocess,
the photon can also be
produced through the {\it fragmentation} of a QCD parton, and the evaluation of the ensuing contribution to the cross section
requires the knowledge of a non-perturbative photon fragmentation function,
which typically has large uncertainties.
The fragmentation contribution is significantly suppressed by the photon isolation criteria
that are necessarily
applied in hadron-collider experiments in order to suppress the large backgrounds.
The {\it standard cone} isolation, which is usually applied
in the experiments, suppresses a large fraction of the fragmentation component.
The {\it smooth cone} isolation completely suppresses the fragmentation contribution \cite{Frixione:1998jh}, and is used in the following with parameters
$R=0.4$ and $\epsilon=0.5$.

We first consider $Z\gamma$ production \cite{Grazzini:2013bna} and we use the cuts
that are applied by the ATLAS collaboration \cite{Aad:2013izg}.
The default
renormalization ($\mu_R$) and factorization ($\mu_F$) scales are set to
$\mu_R=\mu_F=\mu_0\equiv\sqrt{m_Z^2+(p_T^{\gamma})^2}$.
We require the photon to have a transverse momentum $p_T^\gamma>15$ GeV and pseudorapidity $|\eta^\gamma|<2.37$. The charged leptons are required to have $p_T^l>25$ GeV and $|\eta^l|<2.47$, and their invariant mass $m_{ll}$ must be $m_{ll}>40$ GeV.
We require the separation in rapidity and azimuth $\Delta R$ between
the leptons and the photon to be $\Delta R(l,\gamma)>0.7$.
Jets are reconstructed with the anti-$k_T$ algorithm \cite{Cacciari:2008gp} with radius parameter $D=0.4$. A jet must have $E_T^{\rm jet}>30$ GeV and $|\eta^{\rm jet}|<4.4$. We require the separation $\Delta R$ between the
leptons (photon) and the jets to be $\Delta R(l/\gamma,{\rm jet})>0.3$.
Our results for the corresponding cross sections are $\sigma_{LO}=850.7\pm 0.2$ fb, $\sigma_{NLO}=1226.2\pm 0.4$ fb and $\sigma_{NNLO}=1305\pm 3$ fb.
The NNLO corrections increase the NLO result by $6\%$. The loop-induced $gg$ contribution
amounts to $8\%$ of the full NNLO
correction and thus to less than $1\%$ of $\sigma_{NNLO}$.
The corresponding fiducial cross section measured by ATLAS is $\sigma=1.31\pm 0.02~{\rm (stat)}\pm 0.11~{\rm (syst)} \pm 0.05~{\rm (lumi)}$ pb.
We see that the NNLO effects improve the agreement of the QCD prediction with the data, which, however, still have relatively large uncertainties.

\begin{figure}[ht]
\begin{center}
\begin{tabular}{c}
\includegraphics[width=0.47\textwidth,angle=90]{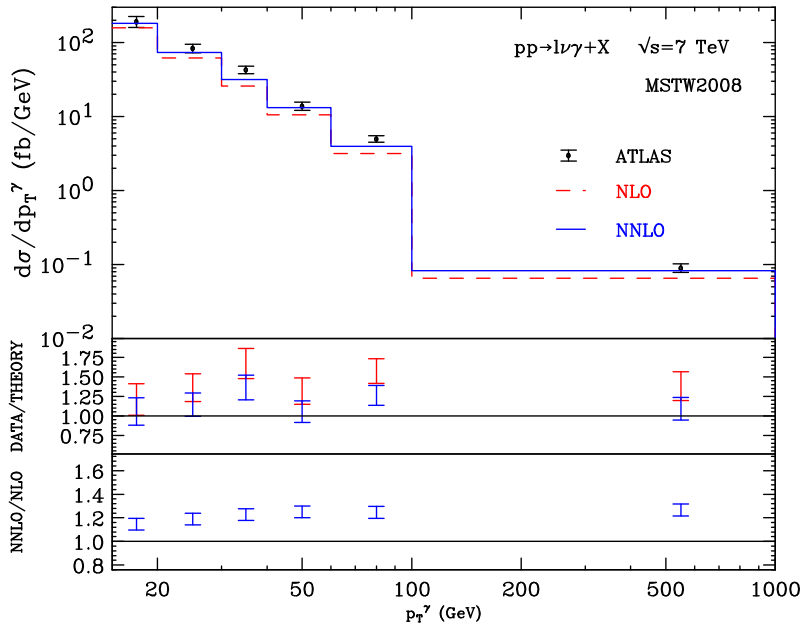}\\
\end{tabular}
\end{center}
\caption{\label{figWg}{The transverse energy distribution of the photon in $W\gamma$ production, computed at NLO (dashes) and NNLO (solid) compared to the ATLAS data. The middle panel shows the ratio DATA/THEORY. The lower panel shows the ratio NNLO/NLO.}}
\end{figure}

We now move to consider $W\gamma$ production \cite{inprep}, and we still use the cuts
that are applied by the ATLAS collaboration \cite{Aad:2013izg}.
The default
renormalization and factorization scales are set to
$\mu_R=\mu_F=\mu_0\equiv\sqrt{m_W^2+(p_T^{\gamma})^2}$.
The cuts are identical to those used for $Z\gamma$ except that the invariant mass cut is replaced by a cut on the missing transverse momentum, $p_T^{\rm miss}$: we require $p_T^{\rm miss}>35$ GeV.
Our results for the corresponding $W\gamma$ cross sections are
$\sigma_{LO}=906.3\pm 0.3$ fb, $\sigma_{NLO}=2065.2\pm 0.9$ fb and $\sigma_{NNLO}=2456\pm 6$ fb. As is well known \cite{Ohnemus:1992jn,Baur:1994sa}, in the case of $W\gamma$ production the QCD radiative corrections are rather large: the NLO corrections increase the LO result by more than a factor of two. The NNLO corrections are thus larger
than those found for $Z\gamma$ and are in fact about $19\%$ for central values of the scales.
The QCD predictions can be compared to the LHC data: the corresponding fiducial cross section measured by the ATLAS collaboration is\cite{Aad:2013izg} $\sigma=2770\pm 30({\rm stat})\pm 330 ({\rm syst})\pm 140 ({\rm lumi})$ fb: we see that the NNLO effect improves the agreement with the data. The same conclusion can be drawn by studying the transverse energy distribution of the photon, as shown in Fig.~\ref{figWg}.

\begin{figure}
\begin{minipage}{\linewidth}
\centerline{\includegraphics[width=0.5\textwidth,angle=90]{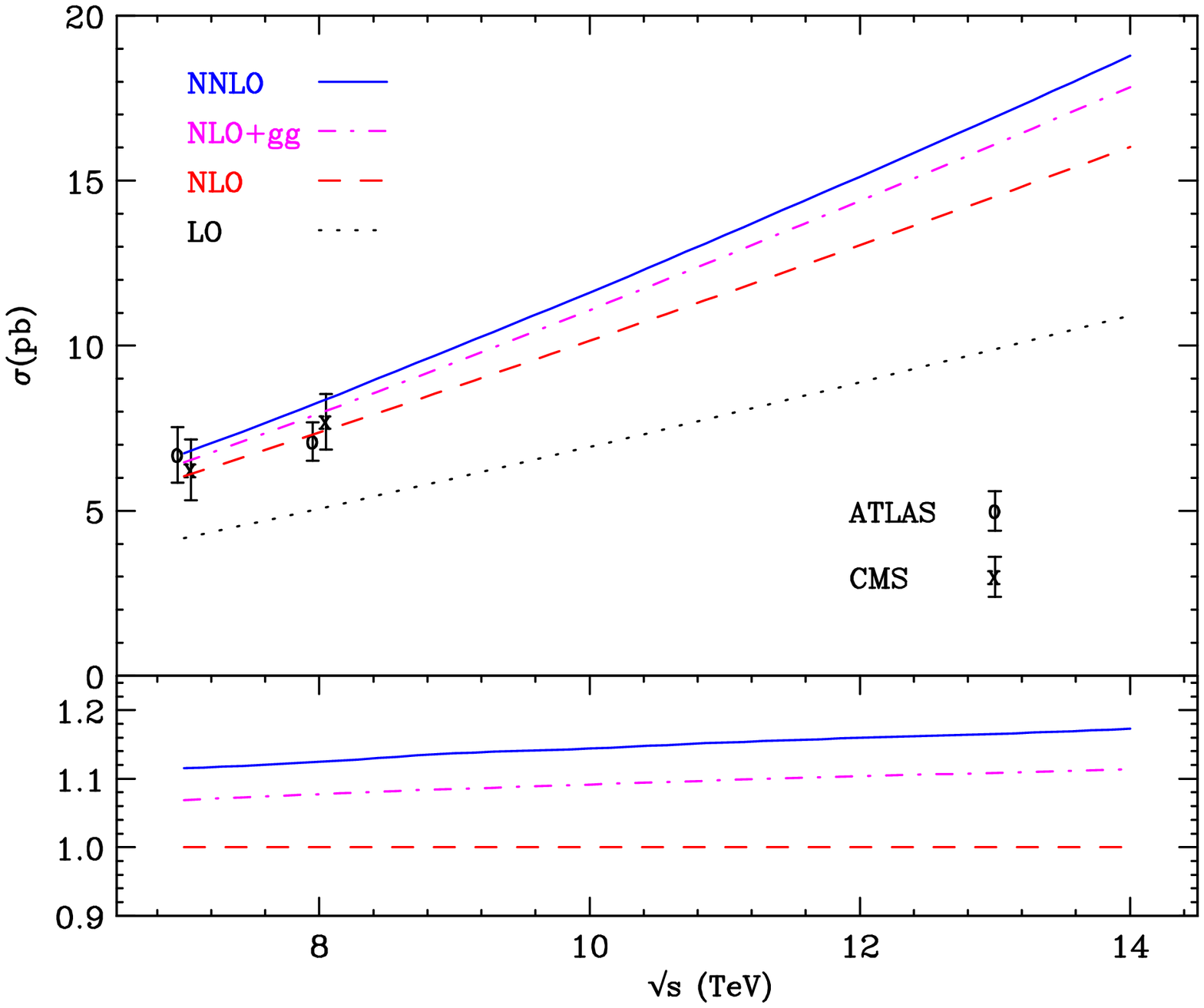}}
\end{minipage}
\caption[]{$ZZ$ cross section at LO (dots), NLO (dashes), NLO+gg (dot dashes) and NNLO (solid) as a function of $\sqrt{s}$. The ATLAS and CMS experimental results at $\sqrt{s}=7$ TeV and $\sqrt{s}=8$ TeV are also shown for comparison \cite{Aad:2012awa,Chatrchyan:2012sga,ATLAS-CONF-2013-020,CMS-PAS-SMP-13-005}. The lower panel shows the NNLO and NLO+gg results normalized to the NLO prediction.}
\label{figZZ}
\end{figure}

We finally present results for the inclusive cross section for $ZZ$ production (see Ref.~\cite{Cascioli:2014yka} for more details).
In this case the default
renormalization and factorization scales are set to
$\mu_R=\mu_F=m_Z$.
In Fig.~\ref{figZZ} we show the cross section computed at
LO, NLO and NNLO as a function of the centre-of-mass energy $\sqrt{s}$.
For comparison, we also show the NLO result supplemented with the loop-induced gluon fusion contribution (``NLO+gg'') computed with NNLO PDFs.
The lower panel in Fig.~1 shows
the NNLO and NLO+gg predictions normalized to the NLO result.
The NLO corrections increase the LO result by about $45\%$. The impact of NNLO corrections with respect to the NLO result ranges
from 11\% ($\sqrt{s}=7$ TeV) to 17\% ($\sqrt{s}=14$ TeV). Using NNLO PDFs, the gluon fusion contribution provides between $58\%$ and $62\%$ of the full NNLO correction.
The theoretical predictions can be compared to the LHC measurements \cite{Aad:2012awa,Chatrchyan:2012sga,ATLAS-CONF-2013-020,CMS-PAS-SMP-13-005} carried out at $\sqrt{s}=7$ TeV and $\sqrt{s}=8$ TeV, which are also shown in the plot.
We see that the experimental uncertainties are still relatively large and that the ATLAS and CMS results are compatible with both the NLO and NNLO predictions. The only exception turns out to be the ATLAS measurement at
$\sqrt{s}=8$ TeV \cite{ATLAS-CONF-2013-020}, which seems to point to a lower cross section.

We have presented a selection of numerical results on $Z\gamma$, $W\gamma$ and $ZZ$ production at the LHC up to NNLO in QCD perturbation theory.
The results for $ZZ$ production were limited to the inclusive cross section for on-shell $ZZ$ pairs. A computation of the two-loop helicity amplitude for $q{\bar q}\to ZZ\to 4l$ will open the possibility of detailed phenomenological studies at NNLO.

\noindent {\bf Acknowledgements.}
I would like to thank Fabio Cascioli, Thomas Gehrmann, Stefan Kallweit, Philipp Maierh\"ofer, Andreas von Manteuffel, Stefano Pozzorini, Dirk Rathlev, Lorenzo Tancredi, Alessandro Torre and Erich Weihs for the nice collaboration on the topics presented in this contribution.
My research was supported in part by the Swiss National Science Foundation (SNF) under contracts CRSII2-141847, 200021-144352,
and by 
the Research Executive Agency (REA) of the European Union under the Grant Agreements PITN--GA---2010-264564 ({\it LHCPhenoNet}), PITN--GA--2012--316704 ({\it HiggsTools}).

\end{document}